# Investigation of the ability to detect electrolyte disorder using PET with positron annihilation lifetime spectroscopy

Radosław Zaleski[1,*], Olga Kotowicz[1], Agnieszka Górska[2], Kamil Zaleski[2] and Bożena Zgardzińska[1]

[1] Maria Curie-Sklodowska University, Institute of Physics, Department of Material Physics,
 Pl. M. Curie-Sklodowskiej 1, 20-031 Lublin, Poland
[2] Medical University of Lublin, Faculty of Medicine, Clinic of Toxicology, Al. Kraśnicka 100, 20-718 Lublin, Poland

[*] radek@zaleski.umcs.pl

**Abstract**

Various concentrations (8÷300 mmol/L) of NaCl, KCl and NaCl + KCl aqueous solutions were investigated using the positron annihilation lifetime spectroscopy (PALS). A strong dependence of the o-Ps intensity as a function of solution concentration was demonstrated. On this basis, the mean positron lifetime or the sum of counts in a selected time interval were proposed as reliable parameters for detecting disturbances in the ion balance of living organisms. The use of these parameters differentiating healthy and cancerous tissues is possible in a new generation of PET scanners equipped with a PALS detection module.

Hyponatremia and hypokalemia are two most frequent electrolyte disorders encountered in cancer patients [1,2]. It has also been established, that concentration of electrolytes is different in cancer tissue than healthy one. Tumorous cells have significantly higher intracellular concentration of sodium and potassium [3]. According to cellular excitability ("CELEX") hypothesis, based on electrophysiological properties of cell membranes, strongly metastatic cancer cells have more electrically excitable membranes which make them aggressive [4]. Overexpression of VGSCs (voltage-gated sodium channels) in many types of cancer causes increase in Na+ influx, which can be linked to mechanisms that are increasing their invasiveness and metastasis [5]. VGSCs are upregulated only in strongly metastatic cells. VGPCs (voltage-gated potassium channels) are downregulated in this phase but play role in proliferation. Thus their activity increases in cells that have reached metastatic sites and grow into secondary tumors. The role of VGSCs is better recognized and can be targeted by anti-metastatic treatment [6]. Higher concentration of sodium in tumor tissue was already measured by magnetic resonance imaging [7,8].

A recently developed total body PET has an ability to measure positron annihilation lifetimes [9-11]. This technology, in addition to the geometric reconstruction of positron annihilation sites, allows to obtain information about the probability of formation and annihilation of positronium – the unstable bound state of an electron and positron. The parameters (lifetime and intensity) of the component in the positron lifetime spectrum, which corresponds to the triplet state of positronium (*ortho*-positronium, *o*-Ps), correlate with the structure of the medium (e.g., the size of the space between molecules can be determined) and its chemical environment, (e.g. it is expected to detect hypoxia [12]). The relationship between the degree of lesions, the type of the investigated tissue and these parameters was discussed [13-15]. The aim of this paper is to investigate if fluctuations in electrolyte concentration is possible to detect with positron annihilation lifetime spectroscopy (PALS).

**Experiment** A digital positron lifetime spectrometer was used in the PALS measurements. It was equipped in an Agilent U1065A digitizer with a sampling rate of 4 GS/s and the software [16] dedicated to the analysis of pulses from two scintillation detectors equipped with $BaF_2$ scintillators. The source of positrons was $^{22}NaCl$ with an activity of 0.5 MBq placed in an 8 μm thick Kapton envelope immersed in a container with a liquid kept at body temperature (309.6 K) degassed using the freeze-pump-thaw technique. For each sample, the positron lifetime spectrum was collected with the number of counts around $1.3 \times 10^7$. The spectra were analysed using PALSfit software. A source correction with an intensity of 9.7% and a lifetime of 382 ps and a resolution function approximated by a single Gaussian with FWHM of 189 ps were assumed. Three components originating from para-positronium (p-Ps), unbound positrons and ortho-positronium (o-Ps) were assumed. This resulted in good fits with $\chi^2 < 1.1$.

**Materials** Ultrahigh purity $H_2O$ (18 MΩ at 298 K) was used to prepare water solutions of NaCl, KCl (POOCH, Poland, p.a.) at concentrations of 8, 16, 25, 50, 75, 150 and 300 mmol/L and their 1:1 mixture at concentrations of 16 and 150 mmol/L.

**Results and Discussion** No significant change with the electrolyte concentration was found in the p-Ps lifetime. Its mean value was (244±6) ps, i.e. with a standard deviation of all results of 6 ps. This value is significantly different from p-Ps lifetime in vacuum of 125 ps. Most likely this is a result of radiolytic processes [17] or formation of the quasi free Ps [18]. The lifetime of unbound positrons shows a small decrease from (500±3) ps to its mean value above 16 mmol/L of (467±8) ps. The mean lifetime of o-Ps ($\tau_{o-Ps}$) is (1.83±1) ns, but possibly the lifetimes for NaCl solutions are slightly smaller (Fig.1). The most pronounced change is in intensities: the o-Ps intensity ($I_{o-Ps}$) decreases from (27.0±0.1)% to (20.3±0.1)% (Fig.1). The $I_{o-Ps}$ strongly depends on the electrolyte concentration and is weakly dependent on the type of water-electrolyte solutions.

This is accompanied by a decrease in p-Ps intensity from (35.2±0.9)% to (25.1±1.2)% and an increase in unbound positrons intensity from (37.9±1.1)% to (54.1±0.8)%. The intensity change with the salt concentration is the same for NaCl and KCl as well as its 1:1 mixtures. This allows to assume that the intensity decreases due to the inhibition of positronium formation by the chloride ions [19]. This implies that PALS will hardly distinguish between sodium and potassium salts but it is sensitive to the concentration of chlorides.

The intensity dependence on the chloride concentration is very well fitted (Fig.1) using the equation:

$$y = \frac{a}{x+b} + c, \qquad (1)$$

where, y is $I_{o-Ps}$ in %, x is the electrolyte concentration in mmol/L, a = (176±20)% × mmol/L, b = (26±3) mmol/L and c = (20.1±0.2)% are the fitting parameters.

Intracellular chloride concentration usually do not exceed 60 mmol/L, which correspond to $I_{o-Ps} > 22$% calculated using Eq.1. The studies of cancerous tissues show that this concentration can increase even 2-3 times [20,21]. This, in turn, corresponds to $I_{o-Ps} < 21$%. This difference is not large, but exceed the standard deviation of $I_{o-Ps}$ ten times.

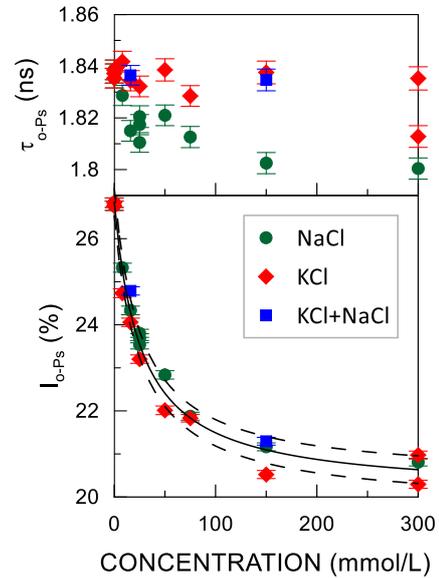

Fig. 1 Ortho-positronium lifetimes and intensities as a function of the concentration of water solutions of NaCl, KCl and their 1:1 mixture. The solid line represents the curve fitted with Eq. 1 and the dashed curves – the confidence interval with a standard deviation.

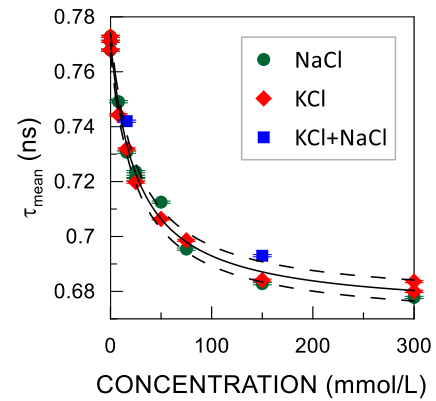

Fig.2 Mean positron lifetime as a function of concentration of water solutions of KCl, NaCl and their 1:1 mixture. The solid line represents the curve fitted with Eq. 1 and the dashed curves – the confidence interval with a standard deviation.

The ratio of the difference between healthy and cancerous tissue to the experimental uncertainty can be even increased using Eq. 1 to fit the dependence of the mean positron lifetime

($\tau_{mean}$) on the chloride concentration (Fig. 2). The $\tau_{mean}$ is given by the equation:

$$\tau_{mean} = \frac{\sum_i \tau_i I_i}{\sum_i I_i}, \quad (2)$$

where $i$ are indices of all positron components in the spectrum. The mean positron lifetime is known to be very reliable parameter that is almost independent on inaccuracies in the fitting (e.g. problems in resolving spectra into components if their lifetimes are too close). Using $\tau_{mean}$ as y in Eq. 1 results in fitting parameters: a = (2.6±0.2) ns × mmol/L, b = (27±2) mmol/L and c = (0.672±0.002) ns. Normal intracellular chloride concentration is indicated by $\tau_{mean}$ > 0.699 ns, while in cancer $\tau_{mean}$ < 0.686 ns. This difference exceeds the standard deviation of $\tau_{mean}$ over 30-times. The scatter of the experimental results exceeding three standard deviations is negligible. Furthermore, a full body scan with PET yields multiple spectra within the cancerous tissue as well as the surrounding healthy tissue. Each of them consists of a high number of counts due to high source activity (~GBq). Therefore, there should be a clear statistical difference in $\tau_{mean}$ due to the variations in chloride concentrations.

Instead of the typical analysis of positron lifetime spectra, only the sum of counts in the 2-12 ns time interval of the spectrum ($N_{2-12ns}$) can be taken into account. This part of the spectrum contains 7-9% of the total counts (Fig.3), which originate almost entirely from the o-Ps component. The contribution of other components above 2 ns is 50 times smaller than the o-Ps component, thus their changes are negligible for $N_{2-12ns}$. There is also a constant contribution from the background of random coincidences, but it does not exceed 0.5%.

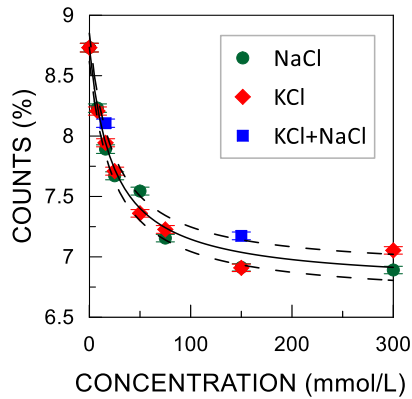

Fig.3 Sum of counts in the 2-12 ns time interval of the positron lifetime spectra as a function of concentration of water solutions of KCl, NaCl and their 1:1 mixture. The solid line represents the curve fitted with Eq. 1 and the dashed curves – the confidence interval with a standard deviation.

This approach does not require any fitting procedure, and therefore allows to obtain uncertainties comparable to the uncertainties of $I_{o-Ps}$ with 16 times smaller number of total counts, i.e. $8 \times 10^6$, in a spectrum. Such number of total counts was collected during one hour. This allowed to study time stability of $I_{o-Ps}$. There were no systematic changes of $N_{2-12ns}$ over time for concentrations up to 150 mmol. Only for the concentration of 300 mmol/L, a slow linear decrease of $N_{2-12ns}$ with a slope of -0.005(2)%/h was observed. This may be due to the accumulation of water radiolysis products over time, which react with $Cl^-$ modifying its inhibition rate. This effect is insignificant for measurement results.

Using number of counts as y in Eq. 1 results in fitting parameters: a = (46±7) mmol/L, b = (23±3) mmol/L and c = (6.77±0.06).

**Conclusions** There is increasing evidence of substantial differences in expression of ion channels and electrolyte concentrations between cancer tissue and healthy ones [1-5]. This provides opportunity to use these differences for both diagnostics and therapeutic purposes under development [6]. The combination of PET imaging with positron annihilation lifetime spectroscopy [9-11] introduces new possibilities for identifying cancerous tissue, e.g. based on the concentration of $Cl^-$ ions. Differences in the $Cl^-$ concentration between healthy tissue and cancer are large enough to be statistically significant in PALS measurements. In particular, the use of the mean positron lifetime or the sum of counts in a selected time interval of the spectrum allows to obtain small uncertainties of the results. Note that these measurements should be from live cells when the chloride transport mechanisms are active.

Additionally, the recent studies suggest synthetic anion transporters can induce apoptosis and can be used in the cancer treatments [22]. The presented method can be used in in vivo studies of such treatment or monitoring its progress.